\documentclass[preprint,12pt]{elsarticle}
\usepackage{enumitem,gensymb}
\usepackage{hyperref}

\usepackage{amsmath,amssymb}
\usepackage{amsthm}

\biboptions{numbers,sort&compress}

\journal{}

\begin{document}

\begin{frontmatter}




\title{Quasienergy operators and generalized squeezed states for systems of trapped ions}
\author{Bogdan M. Mihalcea}
\ead{bogdan.mihalcea@inflpr.ro}
\address{Natl. Inst. for Laser, Plasma and Radiation Physics (INFLPR), \\ Atomi\c stilor Str. Nr. 409, 077125 M\u agurele, Romania}

\begin{abstract}
Collective many-body dynamics for time-dependent quantum Hamiltonian functions is investigated for a dynamical system that exhibits multiple degrees of freedom, in this case a combined (Paul and Penning) trap. Quantum stability is characterized by a discrete quasienergy spectrum, while the quasienergy states are symplectic coherent states. We introduce the generators of the Lie algebra of the symplectic group ${\cal {SL}}(2, \mathbb R)$, which we use to build the coherent states (CS) associated to the system under investigation. The trapped ion is treated as a harmonic oscillator (HO) to which we associate the quantum Hamilton function. We obtain the kinetic and potential energy operators as functions of the Lie algebra generators and supply the expressions for the classical coordinate, momentum, kinetic and potential energy, as well as the total energy. Moreover, we also infer the dispersions for the coordinate and momentum, together with the asymmetry and the flatness parameter for the distribution. The system interaction with laser radiation is also examined for a system of identical two-level atoms. The Hamilton function for the Dicke model is derived. The optical system is modelled as a HO (trapped ion) that undergoes interaction with an external laser field and we use it to engineer a squeezed state of the electromagnetic (EM) field. Such an approach enables one to build CS in a compact and smart manner by use of the group theory. We consider coherent and squeezed states associated to both ion dynamics and to the EM field. 

\end{abstract}


\begin{highlights}
\item Based on the the Bargmann-Fock representation, we show that by \newline applying the time-dependent variational principle (TDVP) for the \newline Schr\"{o}dinger equation on coherent states (CS) we infer the classical Hamilton equations of motion for a single degree of freedom
\item We investigate the time evolution for a multi-ion system and collective many-body dynamics for a combined (Paul and Penning) trap. The solution of the time-dependent Schr\"{o}dinger equation is expressed by means of an evolution operator. In order for the system under investigation to preserve stability, the associated quasienergy spectrum is required to be discrete. Collective variables are employed to characterize the system dynamics, then we introduce the electric potential and identify the equilibrium points (the configurations of minimum) for a 3D quadrupole ion trap (QIT)
\item We demonstrate that time dependent Hamiltonians can be described by means of evolution operators, which can be associated with the symplectic group representation applied in order to build coherent states. The expectation values of the quantum Hamilton function reduced through the evolution operators applied to such states, determine a classical Hamiltonian with a perturbative term that is time periodic
\item We build the coherent states associated to the system by using the generators of the Lie algebra for the symplectic group ${\cal {SL}}(2, \mathbb R)$. 
\item We introduce the Hamilton functions for single-mode and multimode two photon states. Furthermore, the two-photon algebra ${\mathfrak h}_6$ stands for the algebra of the two-photon group ${\cal H}_6$, which is exactly the dynamical group of the unidimensional (1D) oscillator, realized as a semidirect product between the ${\cal N}_2$ group and the ${\cal {SL}}(2, \mathbb R)$ group previously considered. We also prove the test states for the system of interest represent coherent states for the Heisenberg-Weyl ${\cal H}_4$, symplectic ${\cal {SL}}(2, \mathbb R)$, and two-photon groups ${\cal H}_6$, respectively 
\item We introduce the Lie algebra generators for the unitary irreducible representation (UIR) of the symplectic group ${\mathcal S}p\left(2, {\mathbb R}\right)$ and apply this formalism to a harmonic oscillator (HO), for which we derive the expressions for the classical kinetic, potential and total energy as a function of the complex variables used
\item We consider a system consisting of identical two-level atoms which undergoes interaction with laser radiation. We infer the Hamilton function for the Dicke model in case of trapped ions. We model the optical system as a HO interacting with an external laser field and use it to engineer a squeezed state of the electromagnetic (EM field). Such an approach enables one to build CS in an elegant manner, based on the group theory 
\end{highlights}

\begin{keyword}
Paul and Penning trap \sep symplectic coherent states \sep evolution operator \sep quasienergy spectrum \sep phonon \sep quantum control \sep two-photon states \sep Heisenberg-Weyl group

\PACS 02.20.-a \sep 02.20.Sv \sep 03.65.-w \sep 31.15.xh \sep 37.10.Ty 


\end{keyword}

\end{frontmatter}


\section{Introduction}
\label{intro}


The use of group theoretical methods in quantum physics, quantum optics, and generally in theoretical physics, is by now a very well established fact \cite{Brif96, Gerry01, Gil08}. The group-theoretical approach developed by Perelomov \cite{Perel86} and Gilmore \cite{Zha90} represents a versatile tool aimed at building different families of coherent states (CS) \cite{Gua21}. On the other hand, a harmonic oscillator (HO) is a well analysed problem in quantum mechanics \cite{Hus53}. The quantum time-dependent harmonic oscillator (TDHO) has been used as a paradigm to portray the dynamics of many physical systems \cite{Lew69, Comb87a, Gerry87, Comb88, Brown91, Ghe92, Ste92, Major05, Mih09, Davy11, Tiba20a, Zela21a, Meys21, Tiba21}. Quantum dynamics of harmonic oscillators is achieved by means of the so-called {\em Peremolov's generalized coherent states} \cite{Perel86, Buz90} of the Lie algebras \cite{Wei63} associated to the SU(1,1) group and SU(2) group \cite{Brif97, Rosa16, Law18, Gaze21}. Quantum dynamics in a 3D combined ion trap is characterized by the solution of the time-dependent quantum oscillator equations \cite{Ghe92, Major05}. Furthermore, the CS approach leads to quantum solutions that are explicitly constructed as functions of the classical trajectories on the phase space \cite{Comb92}. Coherent states for a set of quadratic Hamiltonians in the ion trap regime are engineered in \cite{Ast10} and then particularized to the asymmetric Penning trap. Dynamical symmetries and the associated symmetry groups have proven to achieve accurate description of trapped ion system dynamics \cite{Ghe92, Ghe00, Major05, Mih11, Schne12, Mih17}. 

The quantum Hamiltonian and the evolution operator can be expressed in terms of the Lie algebra generators, which allows one to obtain the propagator \cite{Iba15}. This approach can be applied to investigate quantum dynamics for charged particles in electromagnetic (EM) fields \cite{Zela20a}, with examples that span linear electrodynamic traps, the Kanai-Caldirola forced HO \cite{Zela21c}, a charged particle in either oscillating magnetic field or constant magnetic field and oscillating electric field (combined Paul and Penning trap). Late investigations demonstrate that by use of the Lewis-Riesenfeld invariant theory and Fock states \cite{Mih09, Law18}, the quantum dynamics of a particle with time-varying mass in a Paul trap can be described. Moreover, the time-dependent Schr\"{o}dinger equation for this problem admits solutions that are used to engineer CS for a quantized particle in the Paul trap \cite{Pedro21}.  

Recent advances in quantum optics enable trapping of single particles or atoms \cite{Bla95, Wine98, Wine13, Sim20, Meys21}, while progress in quantum engineering techniques allows preparing these particles in well-defined quantum states \cite{Vogel01, Sas02, Raja19}, under conditions of accurate control of the interaction between a quantum system (trapped ions) and the environment \cite{Leibf03, Wine11, Lin16, Horn18, Fou19, Mehta20, Wan20}. Besides quantum optics, CS are of large interest starting from pure mathematical topics up to physical applications such as quantum gravity \cite{Berg18, Mar21}. In addition, high precision measurements that include engineering of quantum correlated states, such as quantum tests of physics beyond the Standard Model (SM) or relativistic geodesy \cite{Mehl18, Safro18}, are driven by continuous progress recorded in the development of ultrastable optical atomic clocks \cite{Van15, Koz18}. Late results demonstrate that spin squeezed states, generated using both ultracold atoms and ions, exhibit reduced quantum projection noise which renders them very good candidates for ultraprecise optical clocks \cite{Schu20, Ped20}.  

As methods for direct quantum simulation have been developed with bosons and demonstrated to yield remarkable progress \cite{Ake12, Liu21}, state of the art experimental techniques and methods enable one to extend them to fermionic atoms (such as $^{40}K$, for example \cite{Mod99}) \cite{Casa11, Lama14, Scha17, Xi21} and so study an even wider range of condensed matter systems \cite{Safro18, Koz18}. The $^{88}$Sr$^+$ isotope exhibits advantages similar to those of $^{40}$Ca$^+$ in terms of simplicity of the level scheme and availability of reliable solid-state laser sources required for cooling and interrogation, which makes it an excellent candidate for an optical clock together with other ion species such as $^{27}$Al$^+$, $^{40}$Ca$^+$, $^{115}$In$^+$, $^{171}$Yb$^+$, etc. \cite{Lud15, Van15, Mali21}. Furthermore, the experimental techniques used to average the transition frequency over several Zeeman components in order to suppress the linear Zeeman, electric quadrupole, and quadratic Stark shift have been established and firstly employed on the electric quadrupole reference transition $^2 S_{1/2} \longrightarrow \ ^2 D_{5/2}$ of $^{88}$Sr$^+$ \cite{Dub13}. The use of the odd isotope $^{87}$Sr characterized by a  half-integer nuclear spin $I = 9/2$, has also been explored \cite{Poli13, Van15}. Experimental results show that a high value of the nuclear spin $I$ entails non-negligible problems, such as the case for $^{43}$Ca$^+$ ion \cite{Har14}. 

Cooling mechanisms have been systematically investigated for the $^{88}$Sr atomic species. A flaw of the $^{88}$Sr$^+$ even isotope transition lies in the absence of a hyperfine structure, and hence a magnetic-field-insensitive $m_F = 0 \rightarrow m_F = 0$ transition. An applied magnetic field of a few microTesla ($\mu$T) splits the clock transition into five pairs of Zeeman components, with each pair being distributed symmetrically about the line centre \cite{Poli13}. The atoms belonging to Group 1 and especially Group 2 in the periodical table of elements, provide an abundance of both bosonic and fermionic isotopes. As an outcome of nucleon spin pairing, the bosonic isotopes of the Group 2 (-like) atoms are characterized by even numbered atomic mass and nuclear spin that equals zero. On the other hand, the fermionic isotopes exhibit odd numbered atomic mass and nonzero nuclear spin. The nonzero nuclear spin is associated with hyperfine splitting. This additional complication usually makes bosonic isotopes for somewhat simpler systems for manipulation such as laser cooling. But sometimes a hyperfine structure can bring an unexpected benefit, such as sub-Doppler cooling for fermionic isotopes \cite{Mod99, Ridi11}. For example, the advantage of an even isotope such as $^{172}$Yb$^+$ lies in the absence of a hyperfine structure, which is associated with a low number of electronic states that need to be addressed \cite{Pyka14}. 

We also mention the issue of para-particles and their intriguing nature, as they are neither bosons nor fermions. According to preliminary results para-particles might represent accurate descriptions of physical phenomena like topological phases of matter.  Ref. \cite{Alde21} reports quantum simulation of para-particle oscillators by tailoring the native couplings of two orthogonal motional modes of a trapped ion, where the system reproduces the dynamics of para-bosons and para-fermions of even order in an accurate manner. 

This paper makes use of previous results, which we will shortly review in the current paragraph. It was demonstrated that in case of a bosonic species of nonneutral atoms confined in a Paul trap \cite{Porr08, Lama14}, the quantum equation of motion in the Husimi representation is equivalent with the one describing the perturbed classical oscillator \cite{Mih10a}. The solution of the time-dependent Schr\"{o}dinger equation can be expressed by means of an evolution operator which we build. In order for the trapped ion system to preserve its stability, the associated quasienergy spectrum needs to be discrete, as shown in \cite{Ghe92, Mih11, Mih18}. By applying the time dependent variational principle (TDVP) \cite{Kra81} for the Schr\"{o}dinger equation on CS orbits \cite{Mih17}, one can infer the Hamilton equations of motion on K\"ahler (symplectic) manifolds. Furthermore, we also emphasize that the classical Hamilton functions associated to the system are exactly the expectation values of the quantum Hamiltonian on symplectic CS \cite{Mih11}. The formalism is then applied to Hamiltonians that are nonlinear in the infinitesimal generators of a dynamical symmetry group, which is the case for nonlinear 3D QIT. By using symplectic coherent states the explicit classical equations of motion on the unit disk are derived, for any algebraic model \cite{Wei63} that admits the dynamical group $Sp\left(2, {\mathbb R}\right)$ \cite{Mih18}. It was also proved that the corresponding quasienergy states are explicitly realized as CS parameterized by the stable solutions of the corresponding classical equations of motion. The model is then applied to infer the quantum Hamiltonian for both combined (Paul and Penning) and radiofrequency (RF) nonlinear traps that exhibit axial symmetry. The classical Hamilton functions along with the classical equations of motion are also retrieved for the particular traps investigated \cite{Mih18}. Collective many-body dynamics for time-dependent quantum Hamilton functions was also investigated for a dynamical system that exhibits multiple degrees of freedom, illustrated by means of a combined (Paul and Penning) trap. By applying this analytical model to describe system dynamics, we show that it is straightforward to determine the equilibrium points for a 3D quadrupole ion trap (QIT) \cite{Mih21}.

The paper is organized as follows: Sec. \ref{Gla} reviews some of the fundamental properties of the CS associated to the HO. Based on the Bargmann-Fock representation, we show that by applying the TDVP for the Schr\"{o}dinger equation on CS we infer the classical Hamilton equations of motion for a single degree of freedom. Section \ref{GCS} investigates generalized coherent states (GCS) with an emphasize on particular applications for ion traps. In subsection \ref{phon}  we discuss Coulomb crystals and state of the art techniques used to engineer collective modes of motion for trapped ions, along with their areas of application in quantum sensing and quantum metrology. We also
talk over fault tolerant quantum error correction (QEC) within the frame of QIP schemes. Sec. \ref{Squ} treats squeezed states and ion trap applications. Hamilton functions that characterize single-mode and multimode two-photon systems are discussed in Sec. \ref{Sgl2pho} and \ref{Multi2pho} respectively, as well as the problem of parameterization for squeezed states. We also supply the expectation value of the electric field operator. In Sec. \ref{GSS} we present general squeezed states and introduce the generators of the Lie algebra for the $T_k$ representation of the symplectic group ${\cal {SL}}(2, \mathbb R)$, which we use to build the CS associated to the system. We treat the confined ion as a HO and introduce its associated Hamilton function, then infer the kinetic and potential energy operators as functions of the Lie algebra generators. Finally, we obtain the expressions for the classical coordinate, momentum, kinetic and potential energy, together with the expression of the total energy as a function of the complex variables that we introduce \cite{Mih11, Mih18}. We also derive the dispersions for the coordinate and momentum, along with the asymmetry and flatness parameter for the distribution. Section \ref{laser} investigates interaction with external laser radiation for a system consisting of identical two-level atoms, to which we associate its Hamilton function. We further infer the Hamilton function for the Dicke model using a collective operator to characterize trapped ions. We model the optical system as a HO (trapped ion) that undergoes interaction with an external laser field. Matter light interaction
is also reviewed within the quantum picture frame in subsection \ref{qrm}, where we talk over quantum simulations of the Quantum Rabi Model (QRM) and of the generalized Dicke model in quantum optics. Our approach enables one to elegantly build CS by means of the group theory. Finally, Sec. \ref{Con} is dedicated to a discussion of the results and conclusions. The approach used in the paper is twofold, as we investigate harmonic oscillators (HO) (a) with respect to the modes of the electromagnetic (EM) field and (b) in relation to the quantum dynamics of the trapped ion. While the general formalism applies to both cases, there are circumstances when different approaches are required, therefore the paper treats these situations independently.

\section{Glauber coherent states. Generalized coherent states.}\label{Gla}

We review some key aspects of CS that are relevant for the subject of the paper. A CS represents a minimal uncertainty wave-packet derived by E. Schr\"{o}dinger \cite{Schro26}. Such a state was later introduced to quantum optics by Glauber to represent the states of the quantized electromagnetic (EM) field \cite{Gla07, Choi10}. Glauber has built the eigenstates of the annihilation operator of the HO with an aim to investigate the correlation functions of the EM field \cite{Ortiz19}. At the same time with Glauber and Sudarshan, Klauder created a set of continuous states that included the basic ideas of the CS for arbitrary Lie groups \cite{Kla13}. A decade later, a complete framework to describe the CS associated to Lie groups was achieved, with certain properties that were similar to those of the CS associated to the HO \cite{Gil72, Perel86, Gil08}. The main issue pursued was to achieve a strong connection between CS and the dynamical groups associated with quantum systems, as it is established that the HO system exhibits a dynamical group which is the Weyl-Heisenberg group \cite{Gaze19, Robe21}. Many quantum systems of interest exhibit dynamical symmetry groups to which the generalized Gilmore-Perelomov CS can be associated \cite{Perel86, Buz90, Comb92, Brif96, Kryu13, Ortiz19, Gua21}. These CS determine a framework based on the theory of group representation and on the symplectic geometry of the phase spaces associated as orbits, under which the global dynamical properties of the quantum systems can be discussed. We briefly discussed about the impact and importance of the CS in Section \ref{intro}. 

The linear algebraic models \cite{Wei63} are characterized by an intrinsic property of stability of the CS: if at a certain moment of time the system is described by a CS, it will preserve its state at any subsequent moment of time. The time evolution of the CS follows according to the associated classical trajectories. Furthermore, similar to a classical particle the coherent oscillator states do not spread out with time. From this point of view, the CS provide a natural framework to discuss the relationship between quantum and classical mechanics \cite{Comb92, Nieto00, Land05, Choi10, Mih18}. The classical limit of the quantum correlation functions for systems with many degrees of freedom has been rigorously investigated by Hepp \cite{Hepp74}. Generally, investigations on the occurrence of chaos (both classical and quantum) in quantum systems represent an issue of large interest. Such studies are focused on searching and establishing a correspondence between classical and quantum mechanics \cite{Fla96, Davi16}, based on the CS framework. As the Husimi representation is characterized by classical periodic orbits \cite{Cha86}, it is used to establish connections between the classical trajectories and the wave functions. In addition, in case of classical unstable periodical orbits there exist discontinuities in the Husimi representation \cite{Hell84, Kap99}. The genuine Gaussian generalized coherent states (GCS) for the Hydrogen atom are also called Gaussian-Klauder states. Furthermore, Gaussian-Klauder states associated to the quantized EM field provide a general framework aimed at building Husimi-Wigner distributions, which makes them excellent candidates to establish a correspondence between quantum and classical physics \cite{Choi10}.

The concept of the algebra eigenstates is introduced in \cite{Brif96, Brif97} with an aim to provide a unified description of the generalized coherent states \cite{Phil14, Rosa16} and of the intelligent states associated with a dynamical symmetry group. The formalism is then applied to the two-photon algebra while the corresponding algebra eigenstates are investigated by means of the Fock-Bargmann analytic representation. CS associated with different dynamical symmetry groups exhibit a large interest for quantum physics \cite{Gaze09, Ali14, Anto18}. 

We will consider the Fock space {$\mathcal F_{1}$} for a quantum system that exhibits a single degree of freedom. We denote by $a$ and $a^{\dagger }$ the annihilation and creation bosonic operators, respectively, and by $\left| 0\right\rangle $ the vacuum state vector with unit norm. According to Glauber \cite{Gla07} the coherent states of the EM field are built based on any of the following definitions for the $\left| \alpha \right\rangle $ CS \cite{Leibf03, Major05, Ela21}:

\begin{enumerate}[label=A.\arabic*]
	\item  All CS of the harmonic oscillator may be defined as
	\begin{equation}
		\label{glco14}a\left| \alpha \right\rangle =\alpha \left| \alpha
		\right\rangle \ , 
	\end{equation}
	that is the CS $\left| \alpha \right\rangle$ represent eigenstates of the annihilation operator of the harmonic oscillator (HO), with complex eigenvalues $\alpha$ \cite{Gla07, Dodon02}.
	
	\item  CS result by applying a unitary displacement operator $D\left( \alpha \right) $ on the vacuum state $\left| 0 \right\rangle$ of the HO 
	\begin{equation}
		\label{glco15}\left| \alpha \right\rangle = D\left( \alpha \right) \left| 	0\right\rangle \ ,\; D\left( \alpha \right) = \exp \left( \alpha a^\dagger - \bar \alpha a\right) \ ,
	\end{equation}
	where $\bar \alpha$ represents the complex conjugate of $\alpha$.
	
	\item  CS are quantum states that minimize the uncertainty relationship for the momentum $p$ and coordinate $q$, so that the Heisenberg relationship can be expressed as: 
	\begin{equation}
		\label{glco17}\left( \triangle p\right) \left( \triangle q\right) = \frac \hbar 2\ . 
	\end{equation}
	
	\item  The coherent states are fundamental states associated to the Hamilton function of the HO (with linear terms in $p$ and $q$).
\end{enumerate}

Definition A.$4$ is compliant with Schr\"{o}dinger's example of quantum dynamics by means of CS. Statement A.$2$ illustrates how the Glauber states are coherent states for the Weyl-Heisenberg group \cite{Robe21}, while statement A.$1$ enables one to identify the Glauber states as the intelligent states for this group. Statement A.$3$ characterizes the coherent states as optimal test states to achieve the classical limit. The generalized coherent states (GCS) satisfy one of the definitions given above. The only class of CS that meet all the properties expressed by definitions A.$1$ - A.$4$ are the Glauber states \cite{Gla07}. The standard coherent states (CSs) \cite{Gerry85} have been generalized by Perelomov \cite{Perel86} to more general Lie groups than the Weyl–Heisenberg group. A concise review on the Schr\"{o}dinger-Klauder-Glauber CS, also called canonical CS or standard CS is given in \cite{Gaze09, Ali14}.  
   
The vacuum state vector represents the lowest energy state of the system. The expectation values for the annihilation and creation operators vanish for such state. In addition  
\begin{equation}\label{glco20}
	a\left| 0\right\rangle = 0 = \left\langle 0 \right| a^\dagger \ ,\ \left\langle 0|0\right\rangle = 1 . \  
\end{equation}
The vectors associated to the $n$ boson states can be expressed as 
\begin{equation}\label{glco21}
	\left| n\right\rangle = \left( n!\right)^{-1/2}\left(
	a^{\dagger }\right)^n\left| 0\right\rangle , \ n=0, 1,\ \ldots \ .
\end{equation}
These vectors make up an orthonormal basis in the Fock space $\mathcal F_{1}$. The action of the $a$ and $a^\dagger$ operators on the Fock space basis vectors is given by 
\begin{equation}\label{glco22}
	a\left| n\right\rangle = \sqrt{n}\left| n - 1\right\rangle, \ a^{\dagger }\left| n\right\rangle = \sqrt{n + 1}\left| n + 1\right\rangle ,\  a^{\dagger }a\left| n\right\rangle = n\left| n\right\rangle \ . 
\end{equation}

We use the commutation relations of the creation and annihilation operators in a bosonic system $\left[ a,a^{\dagger }\right] = 1 $, and the Baker-Campbell-Hausdorff (BCH) formula \cite{Wei63, Bial69, Ali14} 
\begin{equation}
	\label{glco41}\exp A\cdot \exp B = \exp \left\{-\frac 12\left[ A, B\right]
	\right\} \exp \left( A + B\right) \ , 
\end{equation}
where $\left[ \left[ A, B\right], A\right] = 0$ and $\left[ \left[ A, B\right], B\right] = 0$. We infer 
\begin{equation} 
	\label{glco42}D\left( \alpha \right) D\left( \beta \right) = \exp \left\{
	\frac 12\left( \alpha \bar \beta - \bar \alpha \beta \right) \right\} D\left(
	\alpha + \beta \right) \ , 
\end{equation}
\begin{equation}
	\label{glco51}\left\langle \alpha |\beta \right\rangle = \exp \left( \bar
	\alpha \beta - \frac 12 \alpha{\bar\alpha} - \frac 12\beta \bar \beta \right) \ ,
\end{equation}
where we have denoted by $\bar \alpha$ and $\bar \beta$ the complex conjugates of $\alpha$ and $\beta$, respectively. The resolution of the unitary operator $\mathbb I$ on the Fock space $\mathcal F_1$ can be expressed as  
\begin{equation}
	\label{glco52}\frac 1\pi \int\limits_{{\mathbb C}}\left| \alpha \right\rangle
	\left\langle \alpha \right| \ d^2\alpha = \mathbb I\ . 
\end{equation}
Out of the relations
\begin{multline}\label{glco54}
D\left( \alpha \right) =\exp \left( -\frac 12\left| \alpha \right| ^2\right)
\exp \left( \alpha a^{\dagger }\right) \exp \left( \bar \alpha a\right) = \\
	\exp \left( \alpha a^{\dagger }-\bar \alpha a\right) = \exp
	\left( \frac 12\left| \alpha \right|^2\right) \exp \left( \bar \alpha
	a\right) \exp \left( \alpha a^{\dagger }\right) \ ,
\end{multline}
we derive the decomposition of the coherent state on the canonical basis of the Fock space, which represents a normalized superposition \cite{Gla07, Ortiz19}: 
\begin{equation}\label{glco56}
	\left| \alpha \right\rangle = \exp \left( - \frac 12\alpha \bar
	\alpha \right) \exp \left( \alpha a^{\dagger }\right) \left| 0\right\rangle
	= \exp \left( - \frac 12\left| \alpha \right|^2\right) \sum_{k = 0}^\infty 
	\frac{\alpha ^k}{k!}\left| k\right\rangle , \ \alpha \in \mathbb C\ . 
\end{equation}

\subsubsection{The Bargmann-Fock representation}

We consider $\left| \psi \right\rangle $ as a normalized unit vector on the Fock space $\mathcal F_1$: 
\begin{equation}\label{glco58}
	\left| \psi \right\rangle = \sum_{n = 0}^\infty c_n\left|
	n\right\rangle ,\ \left\langle \psi |\psi \right\rangle
	=\sum_{n = 0}^\infty \left| c_n\right|^2 = 1\ . 
\end{equation}
We associate a complex function $\Psi$ to the $\left| \psi \right\rangle $ state, defined as  
\begin{equation}\label{glco59}
	\Psi \left( z\right) = \exp \left( \frac 12\left| z\right|^2\right) \left\langle \bar z|\psi \right\rangle = \sum_{n = 0}^\infty
	c_nu_n\left( z\right) ,\ u_n\left( z\right) = \frac{z^n}{\sqrt{n!}}\;. 
\end{equation}
The series defined by eq. (\ref{glco59}) converges uniformly on any compact set of $\mathbb C$. The $u_n$ functions make up an orthonormal basis of the Bargmann space $\mathcal B_1$. The inner product in the Bargmann space is defined as 
\begin{equation}\label{glco63}
	\left( \Psi_1, \Psi_2\right) = \int\limits_{{\mathbb C} }\exp
	\left( -\left| z\right|^2\right) \overline{\Psi}_1\left( z\right) \Psi_2 \left( z\right) d\mu \left( z\right) \ , 
\end{equation}
where the measure $\mu $ is given by $d\mu \left( z\right) = \pi ^{-1}\ dxdy$, with $x = \Re e \ z$ and $y = \Im m \ z$. The Bargmann representation is based on the Fock substitution $a^\dagger \rightarrow z$ and $a\rightarrow d/dz$.

A holomorphic CS $\left| \underline{z}\right\rangle $ can be obtained out of a Glauber state $\left| z\right\rangle $ through 
\begin{equation}\label{g6}
	\left| \underline{z}\right\rangle = \exp \left( \frac 12\ z\bar
	z\right) \left| z\right\rangle \ . 
\end{equation}
In case of a quantum observable $A$, we denote by $A_{cl}\left( z\right) = \left\langle z\right| A\left| z\right\rangle $ the expectation value in the state identified as $\left| z\right\rangle $. In addition, $A_{cl}$ is also a classical observable with respect to the phase space ${\mathcal M} = {\mathbb C}$. The variational principle \cite{Kra81, Mih17} applied for the Schr\"odinger equation on the CS described by eq. (\ref{g6}) determines the Hamilton equations 
\begin{equation}\label{g7}
	i\dot z = \frac{\partial H_{cl}}{\partial \bar z}, \ i{\dot{\bar z}} = - \frac{\partial H_{cl}}{\partial z}\ , 
\end{equation}
where the Hamilton function $H_{cl}$ is associated to the $H$ Hamiltonian on the Fock space $\mathcal F_1$. Ref. \cite{Pli08a} demonstrates that the response properties of a quantum harmonic oscillator (QHO) are in essence those of a classical oscillator. The time evolution of a parametrically driven dissipative quantum oscillator exhibits a strong connection with the classical, damped Mathieu equation \cite{Chak21}, where a close relationship is shown to exist between the localization of the Wigner function and the stability of the damped Mathieu equation. In case of a bosonic species of atoms confined in a nonlinear Paul trap \cite{Mih08}, it was established that the expected value of the quantum Hamiltonian on coherent states results as a function of the eigenvalues of the creation and annihilation operators \cite{Mih10a}. The technique is based on using the Time Dependent Variational Principle (TDVP) \cite{Kra81} followed by deriving the Hamilton equations of motion for a single degree of freedom. Thus, the quantum equation of motion in the Husimi (Q) representation \cite{Harri93} for the bosonic ion system is obtained, which is shown to be equivalent with the equation of motion that describes a perturbed classical oscillator \cite{Mih10a}. Algebraic methods can be used to solve a HO that is initially in the fundamental state, which is then subject of two sudden jumps in the frequency \cite{Tiba20b} or, more generally, for any time-dependent frequency considering it as a sequence of jumps \cite{Tiba20a}. It is also demonstrated that the state of the system after the first jump evolves into a squeezed state of the initial Hamilton function.

It has been also demonstrated that a phase-space representation of quantum mechanics, such as a Husimi-Wigner representation \cite{Ste92, Harri93, Choi10}, reveals the structure of the corresponding phase-space \cite{Meni07}. In particular, for the case of regular classical dynamics, the Husimi function of an eigenstate (or of a Floquet state in the case of a driven system) is localized in the phase-space along the corresponding quantizing torus.

\section{Generalized squeezed states. Exploration of novel physics using ion traps.}\label{GCS}

The dynamics of the Perelomov coherent states \cite{Gerry85} and SU(1,1) squeezing of the SU(1,1) generalized coherent states (GCS) interacting with a nonlinear medium modelled as an anharmonic oscillator, are investigated in \cite{Gerry87, Buz90}. In case of a Paul trap it is demonstrated that the same wavefunctions can either be different coherent or squeezed states \cite{Nieto00}.  The resonant quantum control techniques are susceptible to cumulative errors, particularly if the quantum engineering manipulation techniques demand multiple steps. The resonant control technique is a mechanism that is often used in control theory. Its principle lies in generating the evolution operators by taking advantage of the system sensitivity when interacting with pulses of particular frequencies \cite{Cruz07}. 

As quantum optical models, squeezed states exhibit an essential interest for Quantum Information Processing (QIP) applications \cite{Schu20} and quantum metrology (QM) \cite{Jord20}. Generalized squeezed states exhibit a large interest for quantum optics because in particular cases they supply extra degrees of freedom in the system. Nevertheless, generalized squeezed states can be very difficult to engineer. A recent paper introduces a protocol aimed at generalizing these states, that may be employed to build the squeezed states for any kind of quantum models \cite{Zela18}. The squeezed states linked to the 3D HO can be built as eigenstates of linear contribution of ladder operators which are associated to the generalized Heisenberg algebra \cite{Maz17, Cruz20a}. To achieve optimum control of the quantum harmonic oscillator (QHO), innovative experimental techniques must be explored in order to achieve characterization and suppression of noise. Such techniques are introduced in \cite{Milne21, Kel21} to assess the noise spectrum of a QHO by detecting its response to amplitude modulated periodic drives with a qubit, as they enable evaluation of the intrinsic noise spectrum of an ion trap potential or for atoms in an optical lattice \cite{Xin21}. Multimode macroscopic states consisting of a superposition of spin CS that are generated in a trapped ion system are investigated in \cite{Male18}, while coherent superposition of three motional Fock states of a single trapped ion is investigated in \cite{Corf21}. A late paper investigates quantum dynamics for an ion confined in a RF trap, that undergoes interaction with either a Bose or spin-polarized Fermi gas \cite{Oghi21}. Quantum optical master equations are then derived within the limit of weak coupling and the Lamb-Dicke approximations. 

We also mention investigations towards extending the concept of dynamical maps in case of many-body systems, by considering novel physical phenomena. Of particular interest is the competition between coherent and dissipative multi-particle dynamics, in case of a complex many-body spin model with a universal trapped ion quantum simulator \cite{Schi13}. Laser-cooled and trapped atomic ions represent an almost ideal environment to investigate interacting quantum spin models \cite{Deb18, Lem18}. Ref. \cite{Mon21} reviews theoretical mapping of atomic ions to interacting spin systems, the preparation of complex equilibrium states, and the study of dynamical processes in such many-body interacting quantum systems. Based on these promising results, it is estimated that use of such quantum simulators to investigate complex many-body spin models may expand current knowledge and assure remarkable progress towards explaining the intrinsic features of exotic quantum materials, while also revealing novel phenomena with respect to  the behaviour of interacting quantum systems that cannot be modelled with classical computers. In addition, a 1D chain of trapped ions can be engineered to produce a quantum mechanical system with discrete scale invariance and fractal-like time dependence \cite{Lee19}.

Trapped ions represent an example of a pristine environment that enables scientists to acquire excellent quantum control \cite{Leibf03, Haff08, Sinc11, Wine11, Kau20, War20, Wan20}, which makes them perfect candidates for implementing qudit-based QIP \cite{Bruz19, Low20, Mehta20}. Previous work has not fully explored the practicality of implementing trapped-ion qudits accounting for known experimental error sources. Ref. \cite{Low20} introduces a universal set of protocols for state preparation, single-qudit gates, a generalization of the M\o{}lmer-S\o{}rensen gate for two-qudit gates, and a measurement scheme which achieves shelving to a metastable state. By performing numerical simulations on the sources of error identified from previous trapped-ion experiments, it is demonstrated that there are no limitations to achieve fidelities above 99\% for three-level qudits encoded in $^{137}$Ba$^+$ ions. These methods can be applied to higher-dimensional qudits, such as five-level qudits. For example, quantum entangled qudits in a trapped three-level ion are investigated in \cite{Derm21}, which demonstrates that for entangled qudits, such as qubit-qutrit and qutrit-quadrit, the sudden birth of entanglement can be tuned by the Lamb-Dicke Parameter (LDP). Moreover, as demonstrated in \cite{Corf21}, the motion of a single trapped ion can be controlled using first-order sideband pulses to engineer nonclassical states that exhibit multilevel coherence. To verify this fact it suffices to perform measurements only to the associated qubit system, based only on first-order coupling transitions. As the measurement procedure is quite elementary its applicability also includes any set of entangled qudits, where in particular cases the system under examination can not be accessed by means of interrogating fields.         

Late progress and results illustrate that universal quantum computing (QC) can be generalized to multilevel qudits \cite{Hor19, Der21}. As an outcome of its intrinsic multilevel nature (it can be realized in adequate $d$-level quantum systems), the qudit provides a larger state space to store and process information and the ability to perform multiple control operations simultaneously. The qudit-based QC system can be implemented on various physical platforms, among which ion traps. 

\subsection{Coulomb crystals. Engineering collective modes of motion (phonons) for trapped ions. Quantum metrology and quantum sensing}\label{phon}

Novel ion traps that provide either a static or a dynamic magnetic gradient field allow employing RF radiation to achieve coupling of the internal and motional (external) states of ions, which represents a critical requirement towards achieving conditional quantum logic \cite{Leibf03}. The paper of W\"{o}lk {\em et al} \cite{Wolk17} demonstrates that the Hamiltonian which characterizes coupling in the presence of a resonant dynamic gradient, coincides in a dressed state basis with the system Hamiltonian in case of a static gradient. The coupling strength is characterized in both circumstances by the same effective Lamb-Dicke parameter. Such an approach simplifies things in implementing QIP with trapped ions.

Trapped ions offer an untarnished environment to implement several QC and quantum simulation (QS) \cite{Geo13}  schemes, but the most delicate issue is linked to improving their coherence and scalability. Late experimental progress is focused on novel strategies that boost the coherent interactions in trapped ion systems by employing parametric amplification (PA) of the ions’ motion — based on squeezing the collective motional modes (phonons), which significantly enhances the spin-spin interactions they mediate \cite{Ge19}. In fact, PA enhances quantum spin squeezing (QSS). Moreover, QSS characterizes the reduction of spin noise in a collective spin system and it is important for both entanglement detection and precision metrology  \cite{Wine11, Male18}. The technique opens new pathways towards other applications, such as: (i) employing PA in conjunction with dynamical controls over the driving laser to further suppress unwanted spin-motion entanglement in two-qubit gates, (ii) use stroboscopic parametric driving protocols to augment the amplification of spin-spin interactions.

Ref. \cite{Wolf19} introduces two measurement schemes that rely on a trapped ion prepared in a motional Fock state for displacement and frequency metrology. These schemes are insensitive with respect to the relative phase between the measured interaction and the non-classical quantum state, an issue that requires extra precautions in case of squeezed or Schr\"{o}dinger-cat states. The technique provides a sensible reduction of the quantum noise below the standard quantum limit set by quantum vacuum fluctuations, which considerably enhances the achievable sensitivity of quantum sensors \cite{Jord20}.

An alternative approach is tested in \cite{Dre20}, where it is demonstrated that the motion of a cold trapped ion can be squeezed by modulating the intensity of a phase-stable optical lattice placed inside the ion trap. The method suggested is both reversible (unitary) and state selective. Fully controllable and re-configurable quantum lattices are explored in \cite{War20}, with individually trapped ions in multi-dimensional configurations. In an iconic paper \cite{Meek96}, the creation of a squeezed state of motion was demonstrated by irradiating an ion with a pair of Raman beams. The ion was initially laser cooled to the quantum ground state. A similar technique was later used \cite{Kie15} to engineer and discuss the properties of squeezed states. Such methods rely on the idea that squeezing occurs when an atom is located within a potential that is modulated at twice the trapping frequency.  One can use either {\em travelling standing waves} or dissipation \cite{Chak21} to achieve squeezing. 

In a recent approach \cite{Burd19} squeezing is achieved by means of a temporal modulation of the trapping potential, where the technique is reversible. Besides, squeezing induced by a certain modulation can be reversed by applying a second, temporal driving. The technique introduced in \cite{Dre20} is an improved one, as it places the ion in a valley (crest) of an optical lattice (OL) through a time-varying intensity. Hence, a time-varying potential is created that depends on the ion internal (electronic)  state, which allows the experimenter to perform squeezing of the ion dynamics in a state selective way. Furthermore, the technique can be employed to engineer a {\em control-squeeze} (CSq) gate. Thus, a method to create state dependent squeezing is demonstrated and explored in the paper. Applications include engineering of special quantum states for metrology. We mention Ref. \cite{McCor19},  where enhanced quantum sensitivity is demonstrated for an equal superposition of two eigenstates with maximally different energies. Such a state ideally achieves the peak interferometric sensitivity allowed by quantum mechanics. The technique enables one to generate superpositions of a HO ground state and a number state of the form $\frac 1{\sqrt 2} \left( |0\rangle + |n\rangle\right)$, with $n$ up to 18 in the dynamics of a single trapped ion. Enhanced sensitivity was observed with respect to changes in the frequency of the mechanical oscillator. This approach is also expected to provide improved characterization of motional decoherence, which represents a fundamental source of error in QIP operations with trapped ions. Along with other experimental approaches mentioned in this section, the technique can be employed to characterize other HO in the quantum regime.  Hence, the method opens new perspectives for quantum metrology (QM) and improved fault tolerant QIP \cite{Parra21}, where HO coherence and system scalability represent severe limitations for current technologies. 

For example, Ref. \cite{Wang20} uses Raman beams to coherently manipulate ion qubits encoded in a 2D crystal consisting of dozens of $^{171}$Yb$^+$ ions, while also investigating heating of the vibrational mode (phonons) in case of a single ion. The quantum engineering technique used to prepare the ground-state of radiant vibrational modes is based on Raman-sideband cooling, followed by an average measurement of the phonon number for the state of interest. A very similar experiment is described in Ref. \cite{Dono21}, that reports characterization and coherent control of Coulomb crystals that exhibit {\em radial-2D} orientation. Similar to \cite{Wang20}, the Coulomb crystals consist of up to 19 $^{171}$Yb$^+$ ions. The intriguing feature of a radial-2D phase crystal lies in the fact that the ion plane superposes with the trap radial plane. In addition, the paper studies micromotion effects by measuring the axial (transverse) vibrational mode (phonon) spectrum for a 7-ion crystal deep in the radial-2D regime. Comparable to Ref. \cite{Wang20}, two Raman transitions are employed to achieve spin-motion coupling and coherent excitation of the crystal modes. A careful choice of the wavevector difference in the Raman beams results in strong coupling to the axial modes and suppression of coupling to the radial (or in-line) modes. To characterize the eventual heat transfer between radial and axial crystal directions, heating rate measurements are performed  for the axial center-of-mass (COM) mode by means of resolved sideband spectroscopy. Using identical experimental conditions the heating rates were measured for both a single ion and a 7 ion radial-2D crystal.  

An experiment aimed at improving both coherence and scalability reports on separately addressed entangling gates in chains consisting of up to 25 trapped atomic ions \cite{Ceti22}.  When individual optical addressing is achieved, 
a single long chain of trapped ions provides for fully coupled and reconfigurable quantum gate operations.  Ion axial motion (confinement is frail along this direction) can result in damaging their coupling with the narrow laser beam, which represents a restricting source of control noise. The effect is that the ion chain is expanded and noisy electric fields induce undesirable heating. Ref. \cite{Ceti22} supplies a solution aimed at inhibiting this noise source by performing sympathetic cooling of the qubits in the ion chain. The Raman beams induce Rabi oscillations between the ground hyperfine clock states of a single trapped $^{171}$Yb$^+$ ion.  Axial heating is investigated for both a single ion, as well as for chains of 15 and 25 ions. Because axial confinement declines with the ion chain length, it is estimated that higher cooling powers per ion are required to  compensate heating due to electric field induced noise. The solution lies in employing ion traps operating under cryogenic conditions. Ion chain confinement in optical lattices or in optical tweezers is expected to enable expanding the number of ions to hundreds of them.  

As already emphasized, one of the biggest challenges lies in scaling up and parallelizing quantum computations with long 1D ion strings, as the ion modes of motion favour qubit-qubit coupling mechanisms. Nevertheless, novel methods are devised and tested that focus on implementing scalable and parallel entangling gates, by employing engineered localized phonon modes. Such localized modes of motion are customized by
tuning the local potential of individual ions with programmable optical tweezers \cite{Ols20}.  Hence, all the experimental approaches reviewed in this sub-section establish ion Coulomb crystals levitated in Paul traps as very promising candidates to implement different schemes in QC, QM and quantum sensing.

\subsection{Evolution operators for levitated trapped ion systems}

Quantum dynamics for time-dependent Hamiltonians has been investigated by using multiple methods and techniques \cite{Lew69, Cook85, Comb87a, Comb88, Comb90, Brown91, Ghe92, Ste92, Ghe00, Leibf03, Ber00, Wolk17, Tiba20c}. An analytical model that aims at establishing a framework used to address the issue of the time-evolution of a two-level system is given in \cite{Enri17}. The Hamiltonian is expressed in terms of the Hubbard operators, which makes it easier to characterize the algebraic properties of the observables of interest, while also allowing one to extend the validity of the results for multipartite systems. In addition, the time-evolution operator can be expressed in the disentangling form, which means that the time evolution of the two-level system can be converted into a system of non-linear differential equations. This approach is consistent with the Wei-Norman theorem \cite{Wei63}, that was particularly used in case of the SU(2) and SU(1,1) Lie groups. Such formalism is then used to investigate the problem of a two-level atom interacting with a circularly polarized field \cite{Enri17} or to discuss the nuclear magnetic resonance phenomenon \cite{Enri18}.  

Herein we  investigate the time evolution of a system consisting of $N$ identical ions of mass $M$, confined in a 3D quadrupole ion trap (QIT) that exhibits axial (cylindrical) symmetry \cite{Mih18}. The trap is also characterized by a constant axial magnetic field and a time varying quadrupole electric potential, which is typical for a combined (Paul and Penning) trap. The residual interaction can be treated as a perturbation. We further consider a dynamical system with $n$ degrees of freedom, described by the quantum Hamilton function

\begin{equation}\label{ber1}
H = \sum_{k = 1}^n \left( - \frac{\hbar^2}{2m} \frac{\partial^2}{\partial
x_k^2} + \lambda \frac m2 x_k^2 \right) + V \left( \vec {\textbf x} \right) +
c \left( \vec {\textbf x} \cdot {\vec {\textbf p}} + \vec {\textbf p} \cdot {\vec {\textbf x}} \right) \ .
\end{equation}

Both the axial $\left( n = N\right) $ and radial $\left( n = 2N\right) $ dynamics of the ion system are characterized by the Hamilton function given by eq. (\ref{ber1}). We have denoted $\vec{\textbf x} = \left(x_1, x_2, \ldots, x_n\right)$ and $\vec{\textbf p} = \left(p_1, p_2, \ldots, p_n\right)$, where $p_k = -i \hbar \partial /\partial x_k$ stands for the momentum operator that corresponds to the coordinate $x_k$. The control parameters $\lambda $ and $c$ are time periodic, while $V$ represents a homogeneous potential that is invariant with respect to translations of order $-2$ \cite{Major05, Mih11, Mih18}. The issue of the physical interpretation of these control parameters for Paul, Penning and combined traps is explicitly treated in Ref. \cite{Major05}, as well as in Ref. \cite{Mih17}, and we consider it to be outside the aim of this paper. Nevertheless, we will supply the expression for the quantum Hamiltonian in case of a  particle of mass $m$ and electric charge $Q$, trapped within a magnetic field, homogeneous along the axial direction $x_{3}$, and superimposed over an electric field obtained from the harmonic potential \cite{Mih17}

\begin{equation}\label{w7}
	V = \sum_{i,j = 1}^{3}\zeta_{ij}x_{i}x_{j}\ ,  
\end{equation}
with $\zeta_{ij}$ constant coefficients or time periodic functions which satisfy $\zeta_{11} + \zeta_{22} + \zeta_{33} = 0$. The corresponding quantum Hamilton function can be expressed as 

\begin{equation}\label{w8}
	H = \frac 1{2m}\sum_{j = 1}^3 p_j^2 + QV + \frac m8 \ \omega_c^{2}(x_1^2 + x_2^2)-\frac {\omega_c}2(x_1p_2 - x_2p_1)\ ,
\end{equation}
where $\omega_c = QB_0/m$ denotes the cyclotron frequency for a Penning trap, $B_0$ is the homogeneous axial magnetic field, while the momentum operators are  $p_j = -i\hbar \partial /\partial x$, $1\leq j\leq 3$. We now turn to the quantum Hamilton function for a charged particle confined within a 3D QIT that exhibits cylindrical symmetry 
\begin{equation}
	H_{2} = \frac 1{2m}\left(-i\hbar \vec{\nabla}-\frac Q2 \vec B\times \vec r\right)^2 + QA(t)\left(x_1^2 + x_2^2 - 2x_3^2\right) \  ,  \label{w9}
\end{equation}
where $\vec r = (x_1,x_2, x_3)$ stands for the position operator. In case of a Penning trap $A$ is a constant \cite{Major05, Kno15}. For an electrodynamic (Paul) trap, $A\left(t\right)$ is a time periodic function, of period $2\pi /\Omega $, where $\Omega$ represents the RF voltage frequency. In particular, the Paul trap is characterized by the absence of the magnetic field $B_0 = 0$, and 
\begin{equation}
	A(t)=(r_0^2 + 2z_0^2)^{-1}\left(U_0 + V_0\cos \Omega t\right) ,
	\label{w10}
\end{equation}
where $r_0$ and $z_0$ represent the radial and axial semi-axes of the 3D ion trap electrodes, while $U_0$ and $V_0$ stand for the d.c. and RF trapping voltages.

We now revert to the solution of the time dependent Schr\"{o}dinger equation, which can be expressed as \cite{Comb88, Major05}

\begin{equation}\label{ber2}
	\Phi \left( t\right) = U\left(t, t_0\right) \Phi \left( t_0\right) \ ,
\end{equation}
where $U\left(t,  t_0\right) $ is a unitary evolution operator that can be explicitly built as  
\begin{equation}\label{ber3} 
	U\left( t, t_0\right) = S\left( t\right) \exp \left[-i\left( \tau -\tau_0\right) H_0\right] S^{-1}\left( t_0\right)\ ,
\end{equation} 
with

\begin{subequations}\label{ber4}
	\begin{eqnarray}
		S\left( t\right) = \exp \left( i\alpha \vec{\textbf z}\ ^2/2\right) \exp \left[ -i\beta \left( \vec{\textbf z}\cdot {\vec{\boldsymbol \nabla }} + {\vec {\boldsymbol \nabla }} \cdot \vec{\textbf z} \right) \right] \ , \\
		H_0 = - \frac 12 {\vec{\boldsymbol \nabla}^2} + m \hbar^{-2} V\left( \sqrt{\hbar /m}\  \vec{\textbf z}\right) + \frac 12 \vec{\textbf z}^2\ , \\
		\vec{\textbf z} = \sqrt{m/\hbar }\ e^{- \alpha }\vec{\textbf x}\;,\ \vec{\boldsymbol \nabla} = \left(\partial/\partial z_1, \ldots, \partial /\partial z_n\right) \;.	
	\end{eqnarray}
\end{subequations}

The time dependent functions $\alpha$, $\beta$ and $\tau $, result out of the Hill equation
\begin{equation}\label{ber5}
	\ddot \zeta + \left(\lambda - 2 \dot c - 4 c^2\right) \zeta = 0 \ ,
\end{equation}
with $\zeta = \exp \left( \alpha + i \tau \right),\; 2\beta = \dot \alpha - 4c$. The case of a 3D QIT $\left(c = 0\ ,\; V = 0\right)$ is investigated in \cite{Comb87a}. The stability regions characteristic to this equation determine the control parameters for stable classic motion. Quantum stability in a 3D ideal QIT is characterized by a discrete quasienergy spectrum \cite{Ghe92, Major05, Mih18}
$$
\varepsilon_j= \mu \left(2j + E_0\right) ,\;\;j = 0, 1, \ldots ,
$$
where $\mu E_0$ is the fundamental state quasienergy, and $\mu$ stands for the Floquet exponent that corresponds to the classical stability regions. As demonstrated in \cite{Ghe92, Mih17} the quantum quasienergy states are symplectic CS. We employ the analytical model presented in \cite{Mih21} and introduce the relative coordinates $y_{\alpha j}$ along with a set of collective variables $s$, defined as:

\begin{equation}\label{ber6}
	y_{\alpha j} = x_{\alpha j} - \frac 1N \sum_{\alpha = 1}^N x_{\alpha j,} \ ,\;s = \sum_{j = 1}^d \sum_{\alpha = 1}^N y_{\alpha j}^2 \ ,
\end{equation}
with $x_{\alpha j} = x_{\alpha + j\left(N - 1\right) },\;1\leq \alpha \leq N,\;1\leq j\leq d$. We choose an electric potential that can be expressed as

\begin{equation}\label{ber7}
	W = \frac{bs}2 + 2a\sum_{\mu ,\nu }C_{\mu \nu }V_{\mu \nu }\ , \;
	V_{\mu \nu } = s^{-\mu }\sum_{\alpha \neq \beta }\left( \sum_{j = 1}^d \left(x_{\alpha j} - x_{\beta j}\right)^2 \right)^{\nu - 1} \ .
\end{equation}

Integrable Hamilton functions that exhibit a discrete quasienergy spectrum can be obtained if $C_{\mu \nu } = 0$ for $\mu = \nu$. The Coulomb interaction is defined by $\mu = 0$ and $\nu = 1/2$. The classical Hamilton function is determined by the expectation values of the quantum Hamilton function $H$ on coherent oscillator and symplectic states \cite{Ghe92, Mih11}. The equilibrium points result as solutions of a system of $n = Nd$ equations:
\begin{multline}\label{ber8}
by_{\alpha j} - a\sum_{\mu ,\nu }\mu s^{-1}y_{\alpha j}C_{\mu \nu }V_{\mu \nu} \  + \\
2 a\sum_{\mu, \nu}\left( \nu - 1\right) s^{-\mu }C_{\mu \nu }\sum_{\alpha \neq \beta }\left( x_{\alpha j} - x_{\beta j}\right) \left( \sum_{k = 1}^d \left(x_{\beta k} - x_{\alpha k}\right)^2\right)^{\nu -2} = 0 \ .
\end{multline}

The configurations of minimum are exactly the regions where ion crystals are created \cite{Major05, Li12, Mih18, Mih21}. In case of a one dimensional (1D) integrable dynamical system with $N$ particles, that corresponds to the Calogero potential \cite{Calo71} 

\begin{equation}\label{ber9}
	V = g\sum_{\alpha \neq \beta }\left( \xi _\alpha -\xi _\beta \right)^2, \quad \xi_\alpha = \left( b/ag\right)^{-4} x_{\alpha 1}\ ,
\end{equation}
where the coordinates of the equilibrium configuration $\xi_\alpha $ are exactly the zeros of the Hermite polynomial of order $N$. These solutions establish the minimum points of the potential function, and consequently the ordered structures (ion crystals). A quantum approach for a three ion system is used in \cite{Comb90}. Thus, time dependent Hamilton functions (characteristic for 3D RF traps) can be described by means of evolution operators, which can be associated with the symplectic group representation applied in order to build CS. The expectation values of the quantum Hamilton function reduced through the evolution operators applied to such states, determine a classical Hamiltonian with a perturbative term that is time periodic \cite{Mih18}. By averaging this Hamiltonian \cite{Cas09}, an autonomous dynamical system results whose equilibrium configurations determine the family of ordered structures (ion crystals) \cite{Li12, Mih21}. 

\subsection{Squeezed states}\label{Squ}

\subsubsection{A few words about squeezed states} 

In agreement with the Heisenberg uncertainty principle, the quantum noise inherent in a beam of light limits the amount of information transmitted through the beam \cite{Zhou16}. Any technique which can bend or squeeze the uncertainty circle into an ellipse can in principle be used, with an aim to lower the amount of noise in one of the quadratures. Such squeezing does not violate the uncertainty principle. It rather places the large uncertainty into a quadrature which is not involved in the information transmission process. The technique to squeeze the error ellipse has been firstly explored by Yuen \cite{Yuen76, Yuen83}. The method implies applying a classical source to control the two-photon emission and absorption processes, in a similar manner in which single-photon processes can be employed to generate CS of the radiation field. The states produced by employing this technique were initially named two-photon coherent states. Their properties have been investigated in \cite{Stol70}, with an aim to characterize minimum uncertainty states (packages) \cite{Kryu13}. The same issue was explored by Lu \cite{Lu71, Lu72}, who named them as {\em new coherent states}. The term of squeezed states has been introduced in \cite{Holle79}. 
Coherent and squeezed states for trapped ions have found applications in QIP \cite{Wine98, Mehta20}, QM and quantum sensing \cite{Male18, Fou19, Wolf19, War20, Birr21}, optical clocks \cite{Schu20}, etc.

\subsubsection{Late results and applications}

Multimode squeezed states exhibit a large interest for QIP applications, such as quantum state sharing, quantum teleportation and multipartite entangled states \cite{Male18, Mehta20, Hor19, Yaz08, Mali21}. It was recently demonstrated that the affine quantization of the cosmological dynamics removes the classical singularity and univocally establishes a unitary evolution. Moreover, the semi-classical portrait based on the affine coherent states exhibits a big bounce replacing the big-bang singularity \cite{Berg18}.

The dynamics of a cold trapped ion can exhibit squeezing. The method described in \cite{Dre20} implements a controlled squeeze-gate by modulating the intensity of a phase-stable optical lattice placed inside the trap. Such an approach is of large interest for QIP, as it is demonstrated that the controlled-squeeze (CSq) gate can prepare coherent superpositions of states which are squeezed along complementary quadratures. These states are also extremely interesting for QM.

A sudden change of a harmonic oscillator’s frequency projects a ground state into a squeezed state, a technique that can bypass the time constraint. The technique is employed in Ref. \cite{Xin21} for the purpose of engineering
squeezed states of motion by sudden changes of the harmonic oscillation frequency of atoms placed in an optical lattice (OL). Based on this protocol, rapid quantum amplification of a displacement operator is demonstrated, that can be used for motion detection. This approach is expected to step up speed in quantum gates, while it also enables quantum sensing and QIP in noisy environments.

Ref. \cite{Zela21b} introduces the {\em associated squeezed states}, expressed as a linear superposition of photon-number states with coefficients determined by the associated Hermite polynomials parameterized by two complex numbers (indicating these states are non-classical). Actually, while the squeezed-vacuum consists exclusively of even-photon number states, the new class of squeezed states includes a counterpart that consists only of odd-photon number states, which means that the vacuum state is absent. Taking into account applications such as the use of squeezed states for the detection of gravitational waves, it is expected that advances towards quantum engineering of minimum uncertainty states would lead to progress in applications such as fundamental tests of quantum mechanics and general relativity, or in the Standard Model Extended (SME) \cite{Berg18}, QIP \cite{Bruz19} and QM \cite{McCor19, Jord20}.

\subsubsection{Single-mode two-photon systems}\label{Sgl2pho}

Squeezed coherent states can be achieved by inducing two-photon processes using a classical source \cite{Bish88}. The field Hamiltonian that describes single-mode two-photon systems can be expressed as 

\begin{equation}
	\label{glbfo1}H = \hbar \omega \left(a^{\dagger }a + \frac 12 \right) + \lambda \left( t\right) a^{\dagger 2} + \bar \lambda \left( t\right) a^2\;.
\end{equation}
The terms in the Hamiltonian above represent the generators of the Lie algebra associated to the symplectic group ${\cal {SL}}(2, \mathbb R)$. The squeezed states $\left| \beta \right\rangle $, with $\beta $ a complex parameter, are CS for the dynamical group: 
\begin{equation}
	\label{glbfo22}\left| \beta \right\rangle = S\left( \beta \right) \left| 0\right\rangle \ , \; 
	\left( \beta \right) = \exp \left[ \frac 12 \left( \gamma a^{\dagger 2} - \bar \gamma a^2\right) \right] \ ,
\end{equation}
where $\gamma = \beta \left|\beta \right|^{-1}\arg \tanh \left| \beta \right| $ in case when $\beta \neq 0$, while $\gamma = 0$ when $\beta = 0$. More generally, a larger class of squeezed states can be obtained by considering the Hamilton function  

\begin{equation}
	\label{glbfo2}H = \hbar \omega \left( a^\dagger a + \frac 12 \right) + f_2 \left(t\right) a^{\dagger 2} + \bar{f_2}\left( t\right) a^2 + f_1\left( t\right) a^{\dagger } + \bar{f_1} \left(t\right) a \;. 
\end{equation}

According to eq. (\ref{glbfo2}), the two-photon algebra ${\mathfrak h}_6$ is spanned by six operators: $a^\dagger a$, $a^{\dagger 2}$, $a^2$, $a^\dagger$, $a$, $I$. The ${\mathfrak h}_6$ algebra stands for the algebra of the two-photon group ${\cal H}_6$, that is exactly the dynamical group of the 1D oscillator described by eq. (\ref{glbfo2}), which can be achieved as a semidirect product between the ${\cal N}_2$ group and the ${\cal {SL}}(2, \mathbb R)$ group previously considered. The ${\cal N}_2$ group is a subgroup of the Weyl-Heisenberg group \cite{Robe21, Howe80} ${\cal H}_4$, with a Lie algebra generated by $a$, $a^\dagger $ and $I$. The ${\cal H}_4$ group is a subgroup of the two-photon group ${\cal H}_6$, which represents a subgroup of the symplectic group ${\cal S}p\left( 4, \mathbb R \right)$.

We will further discuss how to use squeezed states with an aim to minimize the uncertainty associated with measurements of EM (laser) fields. To achieve that, we can express the EM field in terms of single-mode creation and annihilation operators. The expectation values of the monomials of these operators then determine the time-dependent mean value of the EM field. Measurement uncertainties are determined by computing the variance of these monomials. The expectation values are estimated for the following families of test states \cite{Gla07, Ortiz19, Gaze19}:

\begin{enumerate}[label=(\roman*)]
\item single-photon Glauber coherent states $\left| \alpha \right\rangle 
= D\left( \alpha \right) \left| 0 \right \rangle \ $,

\item pure squeezed states $\left| \beta \right\rangle = S\left( \beta
\right) \left| 0\right\rangle \ $, 

\item general squeezed states $\left| \alpha, \beta \right \rangle
= D\left( \alpha \right) S\left( \beta \right) \left| 0\right \rangle \ .$
\end{enumerate}

The test states above represent coherent states for the Heisenberg-Weyl ${\cal H}_4$, symplectic ${\cal {SL}}(2, \mathbb R)$ and two-photon groups ${\cal H}_6$, respectively. If the initial Schr\"{o}dinger equation solution for the Hamilton function described by eq. (\ref{glbfo1}) (respectively, by eq. (\ref{glbfo2})) is a pure squeezed state (respectively, a general one), then the solution at a subsequent moment of time is also a pure squeezed (respectively, general) state. Moreover these states are explicitly determined if the time dependence of the $\left| \alpha \right\rangle $ and $\left| \beta \right\rangle $ parameters is explicitly known. Such dependence can be obtained out of the equations of motion determined by the Time-Dependent Variational Principle (TDVP) \cite{Kra81} applied on squeezed states \cite{Mih17}, which are reducible to Riccati equations. In the two-photon representation, explicit results have been obtained in \cite{Yuen76, Yuen83}.   

\subsubsection{Multimode two-photon squeezed coherent states}\label{Multi2pho}

The Hamilton function that describes the preparation of the EM field in a quantum state through single and two-photon processes driven upon by classical sources is

\begin{multline}\label{glbfo24}
H\left( t\right) = \sum_i \hbar \omega_i\left( a_i^{\dagger }a_i + \frac 12 \right) + \sum_{ij}\left[ f_{ij}\left( t\right) a_i^{\dagger }a_j^{\dagger} + H.c.\right]  \\ 
+ \sum_{ij}g_{ij}\left( a_i^{\dagger }a_j+ \frac 12 \delta_{ij}\right) +\sum_i\left[ h_i\left( t\right) a_i^{\dagger } + H.c.\right] \ . 
\end{multline}
where $\mbox{ H.c.}$ stands for the Hermitian conjugate, which means there are additional terms which are the hermitian conjugate of all the preceding terms that have already been written. The dynamical group associated to this Hamiltonian is the two-photon group. As a consequence of the general theorem on CS, a two-photon CS described by such Hamiltonian (in particular the ground state of the electromagnetic field) will evolve into a two-photon coherent state. The most general $n$ mode two-photon CS can be expressed as $\exp \left(R_{ij} a_i^\dagger a_j^\dagger + r_i a_i^\dagger - H.c.\right) \left|0\right\rangle $. There are several ways to parameterize these coherent states. The simplest ones involve successive application of the generalized displacement operator $D\left( \alpha \right) $ and of the $n$ mode squeezing operator $S\left( \beta \right) $ \c, defined as

\begin{equation}\label{glbfo25}
	D\left( \alpha \right) = \exp \left( \alpha_i a_i^\dagger - {\bar{\alpha }}_i a_i\right) \ , \;\;
	S\left( \beta \right) = \exp \frac 12 \left(\beta_{ij} a_i^\dagger a_j^\dagger - {\bar {\beta }}_{ij} a_i a_j\right) \ .
\end{equation}
In fact, the most useful parameterization is 

\begin{equation}
	\label{glbfo26}\left| \alpha \beta \right\rangle \equiv D\left( \alpha
	\right) S\left( \beta \right) \left| 0\right\rangle \ .
\end{equation}

The statistical properties of the electric field operator $E\left(t\right) $ can be computed when the EM field is prepared in the squeezed state $\left| \alpha \beta \right\rangle $. The squeezed state is engineered out of a Hamiltonian similar to eq. (\ref{glbfo2}), with $f_1\left( t\right) = 0,\ f_2\left( t\right) \neq 0$. The properties of the field are investigated after the state has been prepared and the driving terms $f_1\left( t\right), \ f_2\left( t\right) $ have returned to zero. The electric field in the multi-mode case can be expressed as

\begin{equation}
	\label{glbfo27}E\left( t\right) = 2\lambda \sum_i\sqrt{\omega _i}\left(
	a_i e^{-i\omega_i t} + a_i^\dagger e^{i\omega_i t}\right) \ ,
\end{equation}
where we have assumed a small frequency dispersion. The expectation value of the electric field operator is  

\begin{equation}
	\label{glbfo28}\left\langle \alpha \beta \right| E\left| \alpha \beta
	\right\rangle = \left\langle \alpha 0\right| E\left| \alpha 0 \right\rangle
	= 2\lambda \sum_i\sqrt{\omega_i}\left( \alpha_i e^{-i\omega_i t} + {\bar {\alpha_i}}e^{i\omega_i t}\right)\ .
\end{equation}
By expressing this value as a function of the phased or dephased quadrature $\alpha_i = \left(x_i + i y_i\right) $, we infer

\begin{equation}\label{glbfo29}
	\left\langle \alpha \beta \right| E\left| \alpha \beta
	\right\rangle = 2\lambda \sum_i \sqrt{\omega_i}\left(x_i\cos \omega_i t + y_i\ sin \omega_i t\right) \ .
\end{equation}
That is, the centroid of the distribution or the expectation value of the electric field operator evolves in time in the same way as it would for $n$ uncoupled modes.

\section{Generalized squeezed states}\label{GSS}

The generalized squeezed states \cite{Zela18} are represented by the following vectors in the Fock space $\mathcal F_1$:

\begin{equation}\label{co1}
	\left| n, \alpha, z\right\rangle = D \left( \alpha \right) U\left(z\right) \left| n\right\rangle \ , 
\end{equation}
where $\alpha $ and $z$ are complex variables with $\left| z\right| <1$, while $\left| n\right\rangle $ represents a $n$ boson state with $n = 0, 1, \ldots $, given by the eq. (\ref{glco21}). The unitary displacement operators $D\left(
\alpha \right) $ and $U\left( z\right) $ are defined as follows \cite{Gla07, Ortiz19, Robe21, Dodon02}: 
\begin{equation}
	\label{co4}D\left( \alpha \right) = \exp \left(\alpha a^{\dagger} - \bar
	\alpha a\right) \ ,\;\ U\left( z\right) = \exp \left( \zeta K_{+} - \bar \zeta K_{-}\right) \ , 
\end{equation}
with $z=\left( \tanh \left| \zeta \right| \right) \zeta \left| \zeta \right|^{-1}$ for $\zeta \neq 0,\;\zeta \in {\mathbb C}$. The Fock Space $\mathcal F_1$ can be decomposed in the orthogonal sum of subspaces $\mathcal F_1 = \mathcal F_{+} \oplus  \mathcal F_{-}$, where $\mathcal F_{+}$ has the orthonormal base $\left\{ \left| 2p\right\rangle \right\}_{p\in {\mathbb N}}$ (states with an even number of bosons), while $\mathcal F_{-}$ has the orthonormal base $\left\{ \left| 2p + 1 \right\rangle \right\}_{p\in {\mathbb N}}$ (states with an odd number of bosons). We choose 

\begin{equation}\label{co7}
	K_+ = \frac 12 \left( a^\dagger \right) ^2\ ,\; K_- = \frac 12 a^2\ ,\; K_0 = \frac 12 a^\dagger a + \frac 14 \ . 
\end{equation}

The operators defined by eq. (\ref{co7}) are the generators of the Lie algebra for the $T_k$ representation of the symplectic group $\mathcal Sp\left( 2, {\mathbb R}\right) $, of Bargmann index $k$, where $k = \frac 14$ for $\mathcal F_{+}$ and $k = \frac 34$ for $\mathcal F_{-}$. More precisely, $T_k$ is an irreducible unitary representation of the universal covering group of ${\mathcal S}p\left(2, {\mathbb R}\right) $. Further on, we consider all operators have their domain of definition included in the intersection of the domains of definition of the annihilation and creation operators, $a$ and $a^\dagger$, respectively. The displacement operators satisfy 
 
\begin{equation}\label{co8}
	\left( D\left( \alpha \right) \right) ^{\dagger } = D\left( -\alpha
	\right) \ ,\; \left(U\left(z\right) \right)^{\dagger } = U\left(-z\right)\ . 
\end{equation}
We can write
\begin{subequations}\label{co12}
	\begin{eqnarray}
		\left[a, K_+\right] = a^+\ ,\;\ \left[a, K_0\right] = \frac 12 a \ , \\
	\left[a^\dagger, K_0\right] = - \frac 12 a^\dagger\ ,\;\
		\left[ a^\dagger, K_+\right] = \left[a, K_-\right] = 0\ .
	\end{eqnarray}
\end{subequations}
The operators given by eq. (\ref{co7}) together with the operators $a$, $a^\dagger$ and $I$, generate the Lie algebra of the semidirect product between the ${\cal N}_1$ group and the $T_k\left( {\cal S}p\left( 2,{\mathbb R}\right) \right) $ group, where ${\cal N}_1$ stands for the group of all unitary operators $\exp \left( i\varphi \right) D\left( \alpha \right) $, with $\alpha \in {\mathbb C}$ and $\varphi \in {\mathbb R}$. By using the Baker-Campbell-Hausdorff (BCH) formula \cite{Ali14} we infer 

\begin{subequations}\label{co11}
	\begin{eqnarray}
		D\left( - \alpha \right) a D\left( \alpha \right) = a + \alpha \ ,\\
		U\left( - z\right) a U\left( z\right) = \left( 1 - z\bar z\right)^{-1/2}\left( a + z a^\dagger \right) \ .
	\end{eqnarray}
\end{subequations}
We denote
\begin{equation}\label{co13} 
	A_{cl}\left(z, \alpha \right) = \left\langle n, z, \alpha
	|A|n, z, \alpha \right\rangle \ ,\ \hat A = U \left(-z\right) D\left( -\alpha
	\right) AD\left( \alpha \right) U\left( z\right) \ , 
\end{equation}
where $A$ is a polynomial in $a$ and $a^\dagger$. By using eq. (\ref{co1}), eq. (\ref{co13}) can be recast as 
\begin{equation}\label{co14}
	A_{cl}\left(z, \alpha \right) = \left\langle n |\hat A| n \right\rangle \ . 
\end{equation}

Then, we infer the Lie algebra generators expressions as  
\begin{equation}\label{co16}
	\hat a = \left(1 - z \bar z\right)^{-1/2}\left( a + z a^\dagger \right) + \alpha \ , 
\end{equation}

\begin{equation}\label{co17}
	\hat a^\dagger = \left(1 - z \bar z\right)^{-1/2}\left(a^\dagger + {\bar z} a\right) + {\bar \alpha} \ , 
\end{equation}

\begin{multline}\label{co20}
	{\hat K_+} = \left(1 - z \bar z\right)^{-1}\left[ K_{+} + 2{\bar z}K_0 +
	{\overline z}^2 K_-\right] + \\ {\bar \alpha} \left( 1 - z{\overline z}\right)^{-1/2}\left( a^\dagger + {\bar z} a\right) + \frac {{\bar \alpha}^2}{2} \ , 
\end{multline}
\begin{multline}\label{co21}
	{\hat K_-} = \left( 1 - z{\bar z}\right)^{-1}\left[K_- + 2z K_0 + z^2 K_+ \right] + \\ \alpha \left( 1 - z{\bar z}\right)^{-1/2}\left(a + z a^\dagger \right) + \frac {\alpha ^2}{2}\ , 
\end{multline}

\begin{multline}\label{co22}
\hat K_0 = \left(1 - z {\bar z}\right)^{-1}\left[ {\bar z}K_{-} + \left( 1 + z{\bar z}\right) K_0 + z K_{+}\right] + \\  \frac 12 \alpha \left(1 - z\bar z\right)^{-1/2}\left( a^\dagger + {\bar z} a\right) + \frac 12 {\bar \alpha} \left(1 - z{\bar z}\right)^{-1/2} \left(a + z a^\dagger\right) + \frac {\alpha {\bar \alpha}}2 . 
\end{multline}\ 

We further derive 
\begin{subequations}\label{co23}
	\begin{eqnarray}
		\langle n, \alpha, z |a| n, \alpha, z \rangle = \alpha \ ,\; \langle n, \alpha, z | a^\dagger| n, \alpha, z \rangle = \bar{\alpha} \ , \\
		\langle n, \alpha, z | K_0 | n, \alpha, z \rangle = \frac 12 \left( n + \frac 12 \right) \frac{1 + z\bar z}{1 - z\bar z} + \frac{\alpha \bar \alpha }2 \;, \\
		\langle n, \alpha, z | K_+ | n, \alpha, z \rangle = \frac 12 \left( n + \frac 12 \right) \frac{2\bar z}{1 - z{\bar  z}} + \frac{\bar {\alpha^2}}2\;, \\
		\langle n, \alpha, z | K_-| n, \alpha, z \rangle = \frac 12\left( n + \frac 12 \right) \frac {2z}{1 - z{\bar z}} + \frac{\alpha^2}2\;\;,
	\end{eqnarray}
\end{subequations}
where $\alpha $ stands for the Glauber variable and $z$ represents the pure squeezing variable. We will further consider a HO of frequency $\omega $, described by the following Hamilton function 

\begin{equation}
	\label{co24} H = \frac 1{2m}\ p^2 + \frac 12\ m \omega^2 x^2 \ .
\end{equation}

We introduce the momentum and coordinate operators for a 1D HO of mass $m$, defined as 

\begin{equation}\label{co26}
	p = \sqrt{\frac{m\hbar \omega }2} i\left( a^\dagger - a \right) \ ,
	\;\; x = \sqrt{\frac{\hbar}{2m\omega}}\left( a + a^\dagger \right) \ .
\end{equation}
The expressions of the squared position and momentum operators are  
\begin{equation}\label{co29}
	x^2 = \frac \hbar {m\omega }\left( K_+ + K_- + 2K_0\right) \ , \; p^2 = m\hbar \omega \left( 2K_0 -K_+ - K_-\right) \ .
\end{equation}
Then, the kinetic and potential energy of the HO are characterized by the operators 
\begin{equation}\label{co31}
	E_c = \frac 1{2m}\ p^2 = \frac 12\ \hbar \omega \left(2K_0 - K_+ - K_-\right) \ ,
\end{equation}
\begin{equation}\label{co32}
	E_p = \frac 12\ m \omega^2 x^2 = \frac{\hbar \omega }2 \left(2K_0 + K_{+} + K_{-}\right) \ .
\end{equation}
We denote $u = \Re e\ \alpha $ and $v = \Im m\, \alpha $ as the real and imaginary part of $\alpha$ respectively, and introduce $G = \sqrt{\hbar /2m\omega }$. Then, the classical coordinate and classical momentum can be expressed as 

\begin{equation}\label{co33}
	x_{cl} = 2 Gu\ ,\;\ p_{cl} = \sqrt{2m\hbar \omega }\ v \ .
\end{equation}

It is convenient to use the variables $\xi $ and $\eta $ \cite{Mih11, Mih18}, defined as:
\begin{equation}\label{co34}
\xi = \frac{\left( 1 + z\right) \left( 1 + \bar z\right) }{1 - z\bar z}\ ,\;\; \eta = 
\frac{\left(1 - z\right)\left(1 - {\bar z}\right)}{1 - z \bar z}\ .
\end{equation}

We further infer the expressions for the classical kinetic, potential and total energy as
\begin{subequations}\label{co35}
	\begin{eqnarray}
		\left( x^2\right)_{cl} = \frac \hbar {m\omega }\left[ \left(n + \frac 12 \right) \xi + 2u^2\right] ,\; \left( p^2\right)_{cl} = m\hbar
		\omega \left[ \left( n + \frac 12\right) \eta + 2v^2\right] , \\
		\left( E_c\right)_{cl} = \frac{\hbar \omega }2\left[ \left(
		n + \frac 12\right) \xi + 2v^2\right] ,\; \left( E_p\right)_{cl} = \frac{\hbar \omega }2\left[ \left( n + \frac 12\right) \eta + 2 u^2\right] , \\
		E_{cl} = \left( E_c\right)_{cl} + \left( E_p\right)_{cl} = \hbar \omega \left[ \left( n + \frac 12\right) \frac{1 + z{\bar z}}{1 - z{\bar z}} + u^2 + v^2\right] .
	\end{eqnarray}
\end{subequations}
The dispersions of the coordinate and momentum are 
\begin{equation}\label{co37}
	\left( \triangle x\right)^2 = \left( x^2\right)_{cl} - \left(
	x_{cl}\right)^2 = \frac \hbar {m\omega }\left( n + \frac 12 \right) \xi \ ,
\end{equation}
\begin{equation}\label{co38}
	\left( \triangle p \right)^2 = \left( p^2\right)_{cl} - \left(
	p_{cl}\right)^2 = m\hbar \omega \left( n + \frac 12 \right) \eta \ .
\end{equation}
Consequently,  
\begin{equation}\label{ec:co40}
	\triangle x \triangle p = \hbar \left( n + \frac 12 \right) \sqrt{\xi \eta }\ .
\end{equation}
We can express $\hat x^3 = \hat x^2\cdot \hat x$, which allows us to find  $\left( x^3\right)_{cl}$ as
\begin{multline}\label{coco41}
	\left( x^3\right)_{cl} = 2u G\left( x^2\right)_{cl} + \\ 2G^3 \cdot 2u\left(
	1 - z{\bar z}\right)^{-1}\left\langle n \left| \left[ \left( 1 + z\right) a^\dagger
	+ \left( 1 + {\bar z}\right) a\right]^2 \right| n\right\rangle
\end{multline}
where
\begin{multline}
\left\langle n \left|\left[ \left( 1 + z\right) a^{\dagger } + \left( 1 + {\bar z}\right) a\right]^2 \right| n \right\rangle = \\ \left( 1 + z \right) \left( 1 + {\bar z} \right) \left\langle n \left |2a^{\dagger }a + 1 \right|n \right\rangle = \left( 1 + z \right) \left( 1 + {\bar z}\right) \left(2n+1\right)\ , 
\end{multline}
and 
\begin{multline}\label{coco42}
	\left( x^3\right)_{cl} = 4u G^2\left[2 u^2 + \left( n + \frac 12\right) \xi
	+ \left( 2n + 1\right) \xi \right] = \\ 4u G^2\left[ 2u^2 + \frac 32\left( 2n+1\right)	\xi \right] \ .  
\end{multline}
We then introduce the centred moments of order $n$
\begin{equation}
	\label{co49}\mu_n = \left( x^n\right)_{cl} - \left( x_{cl}\right)^n\ ,
\end{equation}
where $\mu _2$ stands for the dispersion, $\sigma = \sqrt{\mu _2}$ is the standard deviation, and $\mu_3$ represents the asymmetry parameter:
\begin{equation}
	\mu_3 = 6u G^2\left( 2n + 1\right) \xi \ .
\end{equation}
The cases when $\mu_3 > 0$ and $\mu_3 < 0$ correspond to left and right asymmetry, respectively. We can also introduce the Pearson asymmetry coefficient defined as   

\begin{equation}\label{co50}
	C_P = \frac{\mu_3^2}{\mu_2^3} = \frac{36 u^2}{G^2\left( 2n + 1\right)
		\xi }\ .
\end{equation}
Moreover, we also introduce $\mu_4$ as the flatness parameter. The distribution can be characterized as flattened, regular or sharp, according as $\mu_4 < 3$, $\mu_4 = 3$ or $\mu_4 > 3$, respectively. The Pearson flatness coefficient is defined as 
\begin{equation}
	\label{co51}C_A = \frac{\mu_4}{\mu_2^2} .
\end{equation}

\section{Interaction with the laser radiation}\label{laser}

We introduce the bosonic annihilation and creation operators, $a_{ij}$ and  $a_{pq}^\dagger$ respectively, which satisfy the following commutation relations:

\begin{equation}\label{}
\left[ a_{ij},\ a_{pq}^\dagger \right] = \delta_{ip}\delta_{jq}\ ,\;\left[a_{ij},\ a_{pq}\right] = \left[a_{ij}^{\dagger },\ a_{pq}^{\dagger}\right] = 0\ ,\;1\leq i,\ j,\ p,\ q\leq n \ , 
\end{equation}
in the Fock space ${\mathcal F_n}$, for a quantum system with $n$ degrees of freedom, where $\delta_{ip}$ and $\delta_{jq}$ stand for the Kronecker delta function. The normalized vacuum state vector is denoted as  $\phi_0$. Then, $a_{ij}\phi_0 = 0$ when $1\leq i,\  j\leq n$.  

We assume the existence of $N$ identical two-level atoms \cite{Gil72, Leibf03} in each of the $n$ possible states of energies $\varepsilon_1 < \varepsilon_2 < \ldots < \varepsilon_n$. The subsystem consists of $N$ atoms that interact with the laser radiation.  By using laser fields, the internal (electronic) levels of trapped ions can be coherently coupled to each other and with the external motional degrees of freedom of the ions. Under conditions of strong trapping, such coupling is equivalent to the Jaynes-Cummings (JC) Hamiltonian \cite{Peder15}. To simplify things we will also consider that a single mode of the EM field interacts with each pair of levels \cite{Soto15}, while the interaction is presumed as near resonant so that $\hbar \omega_{ji}\simeq \varepsilon_j - \varepsilon_i$. For small enough interactions the Hamilton function can be recast as 

\begin{equation}
	\label{op1}{H} = \sum_{1\leq i < j\leq n}\hbar \omega_{ji} a_{ji}^\dagger a_{ji} + \sum_{i = 1}^n\sum_{\alpha = 1}^N\varepsilon_ie_{ii}^{\left( \alpha	\right) } + V_{int}\ .
\end{equation}
We also presume that the transition from state $j$ to state $i$ $\left(j > i\right) $ is associated with the creation of a photon in the near resonant mode. If the element of the dipole matrix associated to this transition is $\lambda_{ji}$, then the interaction term can be expressed as 
\begin{equation}
	\label{op2}V_{int} = \frac 1{\sqrt{N}}\sum \limits_{1\leq i < j\leq n}\sum_{\alpha = 1}^N\left( \lambda_{ji}a_{ji}^\dagger e_{ij}^{\left(	\alpha \right) } + {\bar \lambda}_{ji} a_{ji} e_{ji}^{\left( \alpha \right)}\right) \ .
\end{equation}

Single particle operators $e_{ij}^{\left( \alpha \right) }$ which describe the $\alpha $ particle then commute with the $e_{rs}^{\left(\beta \right) }$ operators which characterize the particle labelled as $\beta $:
\begin{equation}
	\label{op3}\left[ e_{ij}^{\left( \alpha \right) }, \ e_{rs}^{\left( \beta \right) }\right] = \delta _{\alpha \beta }\left( \delta_{jr} e_{is}^{\left(	\alpha \right) } - \delta_{si}e_{rj}^{\left( \alpha \right) }\right) \ .
\end{equation}
When $\alpha = \beta $ fixed, we infer the commutation relationship for the Lie algebra $\mathfrak u(\mit n)$ of the unitary group ${\cal U}\left( n\right) $. An interaction for which the both the field and atomic operators appear as rank one, such as eq. (\ref{op2}), represents a {\em minimal interaction}. We choose the case of the Dicke model (semiclassical approach) where the laser radiation is characterized as a single quantum mode, while matter is described as a set of two-level systems. In order to infer the Dicke Hamiltonian as a special case, we choose $n = 2$ and 
\begin{equation}
	\label{op4}\left\{ 
	\begin{array}{llll}
		a_{21}^{\dagger } = a^\dagger \ ,\;\;e_{21}^{\left( \alpha \right) } = \sigma_\alpha ^{+} = \left( 
		\begin{array}{cc}
			0 & 1 \\ 
			0 & 0
		\end{array}
		\right) &  &  &  \\
		a_{21} = a\ ,\;\;e_{12}^{\left( \alpha \right) }=\sigma _\alpha ^{-} = \left( 
		\begin{array}{cc}
			0 & 0 \\ 
			1 & 0
		\end{array}
		\right)  &  &  &  \\
		e_{22}^{\left( \alpha \right) } - e_{11}^{\left( \alpha \right) } = \sigma_\alpha^z = \left( 
		\begin{array}{cc}
			\frac 12 & 0 \\ 
			0 & - \frac 12
		\end{array}
		\right)
	\end{array}
	\right. \ .
\end{equation}
We also choose
$
\hbar \omega_{21} = \omega ,\; \varepsilon_2 = \varepsilon/2,\;
\varepsilon_1 = - \varepsilon/2,\; \lambda_{21} = \lambda$ and infer the following expression for the Hamilton function:
\begin{equation}
	\label{op5}{H} = \hbar \omega a^\dagger a + \sum_{\alpha = 1}^N\left(
	\varepsilon_2e_{22}^{\left( \alpha \right) } + \varepsilon_1e_{11}^{\left(
		\alpha \right) }\right) + \frac 1{\sqrt N} \sum_{\alpha = 1}^N \left( \lambda a^\dagger \sigma_\alpha^{+} + {\bar{\lambda }}a\sigma_\alpha^{-} \right) \ .
\end{equation}

Within the framework of the Dicke model \cite{Aedo18}, the Hamilton function can be expressed as
 
\begin{equation}
	\label{op6}{H} = \hbar \omega a^\dagger a + \frac \varepsilon 2\left(
	E_{22} - E_{11}\right) + \frac 1{\sqrt N}\left( \lambda a^\dagger E_{12} + {\bar \lambda} a E_{21}\right) \ ,
\end{equation}
where
$$
E=\sum_{\alpha = 1}^N e_{ji}^{\left( \alpha \right) } 
$$
stands for a collective operator that characterizes all trapped ions. The $\lambda $ parameter accounts for the field-ion coupling.  When discussing the dynamical properties of a quantum system, the Hamilton function represents the starting point. This is an outcome of the fact that the Hamiltonian and its associated Hilbert space completely determine the dynamics of the system in quantum mechanics. In quantum optics, the Hamiltonian associated to the system (that characterizes the interaction between the atomic system and the EM field) can be expressed as:

\begin{equation}
	\label{op7}H = \sum\limits_k\ \hbar \omega_ka_k^\dagger a_k + \sum \limits_\alpha \varepsilon \sigma_0^{\alpha} + \sum \limits_{k, \alpha} \gamma_{k\alpha }\left[ \frac{\sigma_+^{\left(\alpha \right)}} {\sqrt N} a_k + \frac{\sigma_-^{\left( \alpha \right) }}{\sqrt N} a_k^\dagger \right] \ ,
\end{equation}
where $\hbar \omega_k$\ represents the energy associated to the $k$ mode of the EM field, while $\gamma_{k\alpha }$ are the coupling coefficients between the $N$ atom system and the EM field. One of the key assumptions made when engineering the Hamilton function given by eq. (\ref{op7}) considers that each of the $N$ atoms labelled with the index $\alpha $ is a two atom system, therefore its dynamic variables are exactly the common spin operators $\left\{ \sigma_0^{\left( \alpha \right) },\ \sigma_+^{\left( \alpha \right) },\ \sigma_-^{\left( \alpha \right) }\right\} $. Normally, it is considered that the strength of the coupling $\gamma_{k\alpha}$ is constant, that is $\gamma_{k\alpha} = \gamma $. If the atomic system is regarded as a classical source (that is the spin operators $\sigma ^{\left( \alpha \right)}$ are treated as $c$ numbers), then eq. (\ref{op7}) can be recast as 

\begin{multline}\label{op8}
H^F = \sum \limits_k \hbar \omega_k a_k^\dagger a_k + \sum \limits_\alpha \left \langle \varepsilon \sigma_0^{\left( \alpha \right) } \right \rangle + \gamma \sum \limits_{k,\ \alpha }\left[\frac{\left\langle \sigma_+^{\left(\alpha \right) }\right\rangle }{\sqrt N} a_k + \frac{\left\langle \sigma_-^{\left( \alpha \right)}\right\rangle }{\sqrt N} a_k^\dagger \right] = \\
\sum\limits_k \hbar \omega_k a_k^{\dagger}a_k + \sum\limits_k \left[ \lambda_k \left( t\right) a_k^{\dagger } + {\bar \lambda}_k\left( t\right) a_k\right] + \text{constant} = \\
\sum\limits_k H_k^F + \text{constant} ,
\end{multline}

where

\begin{equation}\label{op9}
	H_k^F = \hbar \omega_k a_k^{\dagger } a_k + \lambda_k \left(
	t\right) a_k^\dagger + \lambda_k^{*}\left( t\right) a_k = H_0 + H_{\text{inter}}\ .
\end{equation}

In eq. (\ref{op9}) the $H_0$ operator characterizes the free EM field (or the free HO), while $H_{\text{inter}}$ describes the interaction between the EM field and the time dependent external source. Thus, the optical system is modelled as a HO system that undergoes interaction with an external field. Based on the Hamilton function described by eq. (\ref{op9}), one can engineer coherent states of the field in a more compact and elegant manner by employing the group theory. Such method generalizes the concept of CS to the arbitrary Lie groups \cite{Gaze09, Rosa16, Gaze21, Robe21}. At the same moment, the physical interpretations of the CS become more transparent. 

We will now shortly discuss  the case when the spin operators can be considered as c-numbers and explain why this approach is constructive.  A c-number stands for a {\em classical} number and it refers to any number or quantity which is not a quantum operator, that acts upon elements of the Hilbert space of states of a quantum system. Thus, c-numbers denote ordinary complex-valued quantities whose algebra is commutative. Hence, one can regard the operator formalism for quantum mechanics as a generalized version of the probability theory, in which real-valued random variables are represented by self-adjoint operators on a Hilbert space \cite{Whit00}. In addition, complex-valued random variables are represented by normal operators. Normal operators provide the advantage that the spectral theorem is preserved in their case. The normal operators class comprises unitary operators, Hermitian and skew-Hermitian operators, and positive operators, which also applies to the system we have considered. In fact, a c-number is a random variable represented by a scalar multiple of the identity operator, which means that it is practically a random variable whose value is in fact a constant. Thus, it makes sense to compute the expectation value, variance, and higher moments of a c-number relative to a certain state \cite{Ryder96}. 

\subsection{Quantum picture}\label{qrm}

The Quantized Rabi Model (QRM)  is used to characterize the most fundamental light-matter interaction involving
quantized light and quantized matter, and thus achieve a full quantum picture. In general, the QRM is employed to portray the dipolar coupling between a two-level system and a bosonic field mode. It is only recent that an analytical  solution for all coupling regimes was suggested. Experiments are usually performed within the limits of the Jaynes-Cummings (JC) model \cite{Zeng01, Messi03}, where the rotating wave approximation (RWA) is valid. The validity of the RWA model holds when the ratio between the coupling strength and the mode frequency is little. As an outcome of such thing, the JC model achieves an appropriate description of the physical phenomena that occur when a two-level system couples to a bosonic mode of the EM field. Among the systems that are well characterized by the JC model we can enumerate Cavity Quantum Electrodynamics (CQED) and circuit QED (cQED) systems, to which we add trapped ions, where the latter represent a distinct category of strongly coupled Coulomb systems (SCCSs) \cite{Mih19}. The RWA model is no longer valid when light-matter interaction is increasing in strength, such as is the case in the ultrastrong
coupling (USC) and deep strong coupling (DSC) regimes.

Quantum simulation of the quantum Rabi model in all parameter regimes by means of detuned bichromatic sideband excitations of a single trapped ion is investigated in \cite{Peder15}. It is demonstrated that current experimental approaches particularly reproduce can reproduce the USC and DSC regimes of such typical light-matter interaction. In addition, in connection with these dipolar regimes, the paper also explores controlled generation and detection of trapped ions' entangled ground states by means of adiabatic methods.
 
Ref. \cite{Aedo18} focuses on implementing analog quantum simulation of generalized Dicke models in trapped ions. The technique employed integrates bichromatic laser interactions on multiple ions with an aim to perform quantum engineering with respect to all regimes of light-matter coupling in these models, where the light mode is mimicked by an ion motional mode. Numerical simulations help in characterizing the three-qubit Dicke model both in the weak field (WF) regime, where the JC behaviour emerges, as well as in the USC regime, case when the RWA model is not valid. Furthermore, Ref. \cite{Aedo18} also explores simulations of the two-qubit biased Dicke model in the WF and
USC regimes, along with the two-qubit anisotropic Dicke model in both the USC regime and the DSC regime. The remarkable agreement between the mathematical models and the ion system is a strong argument towards testing and implementing these quantum simulations in the laboratory, based on state of the art or emerging technologies. The formalism introduced in \cite{Aedo18} opens new and promising pathways for the quantum simulation of many-spin Dicke models in trapped ions.

The JC Hamiltonian of cavity QED describes the coupling of a two-level atom to a single mode of the (quantized) radiation field. In case of a trapped ion, the coupling of a single two-level atom to the atom’s (harmonic) motion is absolutely similar, where the difference consists in the fact that the HO associated with a single mode of the radiation field in cavity QED is replaced by that of the atom’s motion, as demonstrated in \cite{Wine98}. Furthermore, the paper explores arbitrary, entangled quantum states of trapped ions, along with different techniques and methods aimed at minimizing the motional decoherence for trapped ions. The JC model is successfully employed to characterize quantum evolution of a system, and especially the (nonlinear) vibrionic dynamics of a trapped ion \cite{Vogel95} under conditions that are away from the Lamb-Dicke regime. In \cite{Krumm18} analytical solutions of the quantized-pump field dynamics are compared with the numerical solutions of the classical pump field, for an explicitly time dependent Hamiltonian. Such an approach enables an in-depth study of the electronic (internal) and the motional (external) quantum dynamics of the ion, along with the time evolution of the non-classicality of the quantum state of motion through the Glauber-Sudarshan function. Furthermore, the combination of the JC  and anti-Jaynes Cummings dynamics present in a trapped ion coupled to multiple modes of motion, simultaneously enables one to engineer effective
Hamiltonians that characterize a system analogous to para-Fermi or para-Bose oscillators \cite{Alde21}.

\section{Conclusions}\label{Con}

Collective many-body dynamics for time-dependent quantum Hamilton functions  is investigated for a dynamical system that exhibits multiple degrees of freedom, illustrated by means of a combined (Paul and Penning trap). Time dependent Hamiltonians are characteristic for 3D Paul traps and they can be described by means of evolution operators associated with the symplectic group representation, applied in order to build coherent states. In addition, the expectation values of the quantum Hamilton function reduced through the evolution operators applied to such states, determine a classical Hamiltonian that exhibits a time periodic perturbative term \cite{Mih18}. By averaging this Hamiltonian \cite{Cas09}, an autonomous dynamical system results whose equilibrium configurations determine the family of ordered structures (ion crystals) \cite{Mih21}. In order that  the trapped ion system preserves its stability, the associated quasienergy spectrum is required be discrete \cite{Ghe92, Mih11, Mih18}. We provide the equilibrium points for a 3D QIT and demonstrate that the configurations of minimum are exactly the regions where ion crystals are created.  

We introduce the generators of the Lie algebra of the symplectic group ${\cal {SL}}(2, \mathbb R)$, which we use to build the CSs associated to the system under investigation. The trapped ion is considered a HO to which we associate the quantum Hamilton function, then infer the kinetic and potential energy operators as functions of the Lie algebra generators. We derive the expressions for the classical coordinate, momentum, kinetic and potential energy, along with the total energy, as a function of the complex variables that we use from our previous analytic models \cite{Mih11, Mih18}. Finally, we infer the dispersions for the coordinate and momentum, together with the asymmetry and flatness parameter for the distribution. The system interaction with laser radiation is also examined, for a system of identical two-level atoms, under the Jaynes-Cummings (JC) model. The Hamilton function for the Dicke model \cite{Aedo18} is retrieved and a collective operator is employed to characterize the trapped ion system. The optical system is modelled as a HO (trapped ion) that undergoes interaction with an external laser field, and then used to engineer a squeezed state of the EM field. The model can be extended in the fully quantum picture by engineering (choosing) a CQED JC Hamiltonian. In case of a trapped ion, the coupling of a single two-level atom to the atom’s (harmonic) motion is similar to CQED, with the notable difference that the HO associated with a single mode of the radiation field in CQED is replaced by the one corresponding to the atom’s motion, as already demonstrated in \cite{Wine98}. Hence, the approach we introduce in the paper enables one to build CS in a compact and smart manner by use of the group theory.     

Our model is valid for any 3D or 2D ion trap, as it enables one to identify equilibrium configurations for trapped ions. Applications span areas like ion crystals and QIP with ultracold ions, many-body collective dynamics, high-resolution spectroscopy, quantum sensing and high-precision measurements.

\section{Acknowledgements}

This work was supported by Romanian Ministry of Education and Research, under Romanian National Nucleu Program LAPLAS VI – contract No. 16N/2019. 

B. M. acknowledges help from Daniel Mart\'{\i}nez-Tibaduiza and Oscar Rosas-Ortiz, for pertinent and constructive comments on the manuscript. B. M. is very grateful to the anonymous reviewer(s) for valuable suggestions and comments on the manuscript that enhanced the purpose and impact of the paper.

\bibliographystyle{elsarticle-num} 






\bibliography{GCS}


\end{document}